\documentstyle[aps,epsfig,floats]{revtex}

\begin{document}

\textwidth = 17 cm
\topmargin = -1cm

\title{\begin{flushright}
\begin{minipage}{3cm}
\small
hep-ph/9702280 \\
NIKHEF 97-009 \\
VUTH 97-2 
\end{minipage}
\end{flushright}
QUARK DISTRIBUTION FUNCTIONS IN A DIQUARK SPECTATOR MODEL\footnote{
Talk presented at Diquarks III, October 28-30, 1996, Torino, Italy}}

\author{P.J. Mulders$^{1,2}$, \underline{J. Rodrigues}$^{1,3}$}

\address{$^1$National Institute for Nuclear Physics and High--Energy
Physics (NIKHEF)\\
P.O. Box 41882, NL-1009 DB Amsterdam, the Netherlands\\
$^2$Department of Physics and Astronomy, Free University \\
De Boelelaan 1081, NL-1081 HV Amsterdam, the Netherlands \\
$^3$Instituto Superior T\'{e}cnico \\
Av. Rovisco Pais, 1100 Lisboa, Portugal}


\maketitle

\begin{abstract}
The representation of quark distribution functions in
terms of nonlocal operators is combined with a simple diquark spectator model.
This allows us to estimate these functions for the nucleon
ensuring correct crossing and support properties.
\end{abstract}


\newcommand{\bm}[1]{\mbox{\boldmath $#1$}}
\newcommand{\tv}[1]{{\bm #1}_{T}}
\newcommand{\ptinst}{\,\tv{p}\!\cdot\!\tv{S}\,}
\newcommand{\ktinsht}{\,\tv{k}\!\cdot\!{\bm S}_{hT}\,}
\newcommand{\nnn}{\nonumber\\}
\newcommand{\com}[1]{($\clubsuit$~{\em #1}~!?!)}


\newcommand{\be}{\begin{equation}}
\newcommand{\bea}{\begin{eqnarray}}
\newcommand{\ee}{\end{equation}}
\newcommand{\eea}{\end{eqnarray}}
\newcommand{\ba}{\begin{array}}
\newcommand{\ea}{\end{array}}
\newcommand{\nin}{\noindent}
\newcommand{\nn}{\nonumber}

\newcommand{\tr}{{\rm Tr}}


\newcommand{\half}{{1 \over 2}}
\newcommand{\terco}{{1 \over 3}}
\newcommand{\quarto}{{1 \over 4}}
\newcommand{\raizd}{{1 \over \sqrt{2}}}
\newcommand{\raizt}{{1 \over \sqrt{3}}}


\newcommand{\galpha}{\gamma^\alpha}
\newcommand{\gbeta}{\gamma^\beta}
\newcommand{\ggamma}{\gamma^\gamma}
\newcommand{\gmu}{\gamma^\mu}
\newcommand{\gnu}{\gamma^\nu}
\newcommand{\grho}{\gamma^\rho}
\newcommand{\gsigma}{\gamma^\sigma}

\newcommand{\gbalpha}{\gamma_\alpha}
\newcommand{\gbbeta}{\gamma_\beta}
\newcommand{\gbgamma}{\gamma_\gamma}
\newcommand{\gbmu}{\gamma_\mu}
\newcommand{\gbnu}{\gamma_\nu}
\newcommand{\gbrho}{\gamma_\rho}
\newcommand{\gbsigma}{\gamma_\sigma}


\newcommand{\smn}{\sigma^{\mu \nu}}


\newcommand{\lc}{\epsilon_{\mu \nu \rho \sigma}}
\newcommand{\lcup}{\epsilon^{\mu \nu \rho \sigma}}


\newcommand{\ga}{\gamma_{5}}
\newcommand{\gp}{\gamma^{+}}
\newcommand{\gm}{\gamma^{-}}


\newcommand{\gma}{g^{\mu \alpha}}
\newcommand{\gmb}{g^{\mu \beta}}
\newcommand{\gmg}{g^{\mu \gamma}}
\newcommand{\gmn}{g^{\mu \nu}}
\newcommand{\gmr}{g^{\mu \rho}}
\newcommand{\gms}{g^{\mu \sigma}}

\newcommand{\gna}{g^{\nu \alpha}}
\newcommand{\gnb}{g^{\nu \beta}}
\newcommand{\gng}{g^{\nu \gamma}}
\newcommand{\gnm}{g^{\nu \mu}}
\newcommand{\gnr}{g^{\nu \rho}}
\newcommand{\gns}{g^{\nu \sigma}}


\newcommand{\hma}{g^{\mu \alpha}}
\newcommand{\hmb}{g^{\mu \beta}}
\newcommand{\hmg}{g^{\mu \gamma}}
\newcommand{\hmn}{g^{\mu \nu}}
\newcommand{\hmr}{g^{\mu \rho}}
\newcommand{\hms}{g^{\mu \sigma}}

\newcommand{\hna}{g^{\nu \alpha}}
\newcommand{\hnb}{g^{\nu \beta}}
\newcommand{\hng}{g^{\nu \gamma}}
\newcommand{\hnm}{g^{\nu \mu}}
\newcommand{\hnr}{g^{\nu \rho}}
\newcommand{\hns}{g^{\nu \sigma}}


\newcommand{\dalpha}{\partial_\alpha}
\newcommand{\dbeta}{\partial_\beta}
\newcommand{\dmu}{\partial_\mu}
\newcommand{\dnu}{\partial_\nu}
\newcommand{\drho}{\partial_\rho}
\newcommand{\Dmu}{D_\mu}

\newcommand{\dcalpha}{\partial^\alpha}
\newcommand{\dcbeta}{\partial^\beta}
\newcommand{\dcmu}{\partial^\mu}
\newcommand{\dcnu}{\partial^\nu}
\newcommand{\dcrho}{\partial^\rho}


\def\sla#1{\setbox0=\hbox{$#1$}
   \dimen0=\wd0 \setbox1=\hbox{/} \dimen1=\wd1
   \ifdim\dimen0>\dimen1 \rlap{\hbox to \dimen0{\hfil/\hfil}} #1
   \else  \rlap{\hbox to \dimen1{\hfil$#1$\hfil}} / \fi}


\newcommand{\alp}{\alpha}
\newcommand{\bet}{\beta}
\newcommand{\gamm}{\gamma}
\newcommand{\Gam}{\Gamma}
\newcommand{\Gamm}{\Gamma}
\newcommand{\del}{\delta}
\newcommand{\Del}{\Delta}
\newcommand{\eps}{\epsilon}
\newcommand{\lam}{\lambda}
\newcommand{\Lam}{\Lambda}
\newcommand{\sig}{\sigma}
\newcommand{\Sig}{\Sigma}


\newcommand{\kt}{{\bm k}_T^2}
\newcommand{\ktns}{{\bm k}_T}
\newcommand{\kp}{{{\bm k}'_T}^2}
\newcommand{\kpns}{{\bm k}'_T}
\newcommand{\qt}{{\bm q}_T^2}
\newcommand{\qtns}{{\bm q}_T}
\newcommand{\pt}{{\bm p}_T^2}
\newcommand{\pts}{{\bm p}_T^2}
\newcommand{\ptns}{{\bm p}_T}
\newcommand{\st}{{\bm S}_T^2}
\newcommand{\stns}{{\bm S}_T}
\newcommand{\sht}{{\bm S}_{hT}^2}
\newcommand{\shtns}{{\bm S}_{hT}}
\newcommand{\xpts}{(x,{\bm p}_T^2)}
\newcommand{\xptns}{(x,{\bm p}_T)}
\newcommand{\zkp}{(z,{{\bm k}'_T}^2)}
\newcommand{\zkpns}{(z,{\bm k}'_T)}
\newcommand{\zkt}{(z,{{\bm k}_T}^2)}
\newcommand{\zktns}{(z,{\bm k}_T)}


\newcommand{\cfa}{\Phi^{[\Gamma]} (x,{\bm p}_T)}
\newcommand{\cfgp}{\Phi^{[\gamma^+]} (x,{\bm p}_T)}
\newcommand{\cfgpga}{\Phi^{[\gamma^+ \gamma^5]} (x,{\bm p}_T)}
\newcommand{\cfsa}{\Phi^{[i\sigma^{i+} \gamma^5]} (x,{\bm p}_T)}


\newcommand{\fo}{f_1}
\newcommand{\fox}{f_1(x)}
\newcommand{\foxk}{f_1(x,{\bm k}^2_T)}
\newcommand{\foxp}{f_1(x,{\bm p}^2_T)}

\newcommand{\go}{g_1}
\newcommand{\gol}{g_{1L}}
\newcommand{\got}{g_{1T}}
\newcommand{\goto}{g_{1T}^{(1)}}
\newcommand{\gos}{g_{1s}}
\newcommand{\gox}{g_{1}(x)}
\newcommand{\gotox}{g_{1T}^{(1)}(x)}
\newcommand{\golxk}{g_{1L}(x,{\bm k}^2_T)}
\newcommand{\golxp}{g_{1L}(x,{\bm p}^2_T)}
\newcommand{\gotxk}{g_{1T}(x,{\bm k}^2_T)}
\newcommand{\gotxp}{g_{1T}(x,{\bm p}^2_T)}
\newcommand{\gosxk}{g_{1s}(x,{\bm k}^2_T)}
\newcommand{\gosxp}{g_{1s}(x,{\bm p}^2_T)}
\newcommand{\gsperp}{g_s^\perp}
\newcommand{\gtwo}{g_2}
\newcommand{\gtwox}{g_2(x)}

\newcommand{\ho}{h_{1}}
\newcommand{\hox}{h_{1}(x)}
\newcommand{\hot}{h_{1T}}
\newcommand{\hotxk}{h_{1T}(x,{\bm k}^2_T)}
\newcommand{\hotxp}{h_{1T}(x,{\bm p}^2_T)}
\newcommand{\holperp}{h_{1L}^\perp}
\newcommand{\hotperp}{h_{1T}^\perp}
\newcommand{\hosperp}{h_{1s}^\perp}
\newcommand{\holperpo}{h_{1L}^{\perp (1)}}
\newcommand{\holperpox}{h_{1L}^{\perp (1)}(x)}
\newcommand{\hotperpt}{h_{1T}^{\perp (2)}}
\newcommand{\hotperptx}{h_{1T}^{\perp (2)}(x)}
\newcommand{\holperpxk}{h_{1L}^\perp (x,{\bm k}^2_T)}
\newcommand{\holperpxp}{h_{1L}^\perp (x,{\bm p}^2_T)}
\newcommand{\hotperpxk}{h_{1T}^\perp (x,{\bm k}^2_T)}
\newcommand{\hotperpxp}{h_{1T}^\perp (x,{\bm p}^2_T)}
\newcommand{\hosperpxk}{h_{1s}^\perp (x,{\bm k}^2_T)}
\newcommand{\hosperpxp}{h_{1s}^\perp (x,{\bm p}^2_T)}
\newcommand{\htwo}{h_2}
\newcommand{\htwox}{h_2(x)}

\newcommand{\ex}{e(x)}
\newcommand{\exk}{e(x,{\bm k}^2_T)}
\newcommand{\expp}{e(x,{\bm p}^2_T)}

\newcommand{\fperp}{f^\perp}
\newcommand{\fperpxk}{f^\perp (x,{\bm k}^2_T)}
\newcommand{\fperpxp}{f^\perp (x,{\bm p}^2_T)}

\newcommand{\gprimet}{g'_T}
\newcommand{\gprimetxk}{g'_T (x,{\bm k}^2_T)}
\newcommand{\gprimetxp}{g'_T (x,{\bm p}^2_T)}

\newcommand{\glperp}{g_L^\perp}
\newcommand{\gtperp}{g_T^\perp}
\newcommand{\glperpo}{g_L^{\perp (1)}}
\newcommand{\glperpox}{g_L^{\perp (1)}(x)}
\newcommand{\gtperpt}{g_T^{\perp (2)}}
\newcommand{\gtperptx}{g_T^{\perp (2)}(x)}
\newcommand{\glperpxk}{g_L^\perp (x,{\bm k}^2_T)}
\newcommand{\glperpxp}{g_L^\perp (x,{\bm p}^2_T)}
\newcommand{\gtperpxk}{g_T^\perp (x,{\bm k}^2_T)}
\newcommand{\gtperpxp}{g_T^\perp (x,{\bm p}^2_T)}
\newcommand{\gsperpxk}{g_s^\perp (x,{\bm k}^2_T)}
\newcommand{\gsperpxp}{g_s^\perp (x,{\bm p}^2_T)}

\newcommand{\gt}{g_T}
\newcommand{\gtx}{g_T(x)}
\newcommand{\hl}{h_L}
\newcommand{\hs}{h_s}
\newcommand{\hlx}{h_L(x)}

\newcommand{\htperp}{h_T^\perp}
\newcommand{\htperpxk}{h_T^\perp (x,{\bm k}^2_T)}
\newcommand{\htperpxp}{h_T^\perp (x,{\bm p}^2_T)}
\newcommand{\htt}{h_T}
\newcommand{\htxk}{h_T (x,{\bm k}^2_T)}
\newcommand{\htxp}{h_T (x,{\bm p}^2_T)}
\newcommand{\hlxk}{h_L (x,{\bm k}^2_T)}
\newcommand{\hlxp}{h_L (x,{\bm p}^2_T)}
\newcommand{\hsxk}{h_s (x,{\bm k}^2_T)}
\newcommand{\hsxp}{h_s (x,{\bm p}^2_T)}
\newcommand{\hto}{h_T^{(1)}}
\newcommand{\htox}{h_T^{(1)}(x)}
\newcommand{\htperpo}{h_T^{\perp (1)}}
\newcommand{\htperpox}{h_T^{\perp (1)}(x)}


\newcommand{\bpsi}{\overline{\psi}}
\newcommand{\bpsix}{\bar{\psi}(x)}
\newcommand{\dsdt}{\int \ [d \sig d \tau]}
\newcommand{\dshdth}{\int \ [d \sig_h d \tau_h]}
\newcommand{\dshdthb}{\int \ \{d \sig_h d \tau_h\}}
\newcommand{\hadront}{2MW^{\mu \nu }}
\newcommand{\mh}{M_h}


\section{Introduction}

Quark distribution functions appear in the field-theoretical 
description of hard scattering processes as the parts that
connect the quark and gluon lines to hadrons in the initial state.
These parts are defined through (connected) matrix elements of nonlocal
operators built from quark and gluon fields. The simplest, but most important
ones, are the quark-quark correlation
functions\cite{Soper-77,Collins-Soper-82,Jaffe-83}.
For each quark flavor one can write down

\be
\Phi_{ij} (p, P, S) = \int {d^4 \xi \over (2 \pi)^4} \ e^{-i p \cdot \xi} \
\langle P, S | \bpsi_j(\xi) \psi_i(0) | P, S \rangle,
\label{qcf1}
\ee
The hadron states are characterized by the momentum and spin vectors, 
$\vert P, S\rangle$. The quark momentum is denoted $p$. 
A summation over colors is understood in $\Phi$.

In a particular hard process only certain Dirac projections of the
correlation functions appear. For inclusive lepton-hadron scattering,
where the hard scale $Q$ is set by the spacelike momentum transfer $-q^2$
= $Q^2$, the correlation function $\Phi$ appears in the leading order
result in an expansion in $1/Q$ and the structure functions
can be expressed in quark distribution functions. 

The hard momentum scale $q$ and the hadron momentum $P$ 
define the lightlike directions $n_\pm$. 
In inclusive lepton-hadron scattering they are related to the hadron 
momentum $P$ and the hard momentum $q$ as
\bea
P & = &
{M^2 x_b \over Q \sqrt{2}} \ n_- + {Q \over x_b \sqrt{2}} \ n_+,
\label{parp} \\
q & = & {Q \over \sqrt{2}} \ n_- - {Q \over \sqrt{2}} \ n_+.
\eea

The quark momentum $p$ and the spin vector $S$ are also 
expanded in the lightlike vectors,
\bea
&&p = {x Q \over x_b \sqrt{2}} \ n_+
+ {x_b (p^2 + \pts) \over x Q \sqrt{2}} \ n_-
+ p_T,
\\
&&S = {\lam Q \over M x_B \sqrt{2}} \ n_+ -
{\lam M x_B \over Q \sqrt{2}} \ n_- + S_T.
\eea
Thus $x$ represents the fraction of
the momentum in the $+$ direction carried by the quark inside the hadron.
The lightlike components $a^\pm \equiv a\cdot n_\mp$ are defined with the
help of the vectors $n_\pm$.

It should be noted that there are other contributions in the leading
cross section resulting from soft parts with $A^+$ gluon legs. These can
be absorbed into the correlation function providing the link operator
that renders the definition in Eq.~\ref{qcf1} color gauge invariant. 
Choosing the $A^+ = 0$ gauge in the study of $\Phi$ the
link operator reduces to unity.

Although hard cross sections can be expressed in terms of distribution
and fragmentation functions, these objects cannot be calculated from QCD
because they involve the hadronic bound states. We follow here a 
different route. We investigate the structure of lightcone correlation 
functions (including the effects of transverse separation of the quark fields),
we illustrate the consequences of various constraints and we estimate the
distribution functions in a simple model. Particularly
suitable is a model in which the spectrum of intermediate states that
can be inserted in Eq.~\ref{qcf1} is replaced by one state, 
which if the hadron is a baryon will simply be referred to as a {\em diquark}.


\section{Quark Correlation Functions}

The most general expression for $\Phi$ is \cite{{ralston79},{piet95}}:
\bea \label{cf2}
\Phi(p,P,S) & = & M \,A_{1} +  A_{2}\, \sla P + A_{3} \,\sla p +
M \,A_{6} \,\sla S \,\ga + {A_{7} \over M} \,p \cdot S \ \sla P \,\ga \nn \\
& & + {A_{8} \over M} \,p \cdot S \ \sla p \,\ga 
+ i \,A_{9} \,\smn \,\ga \ S_{\mu} P_{\nu} \nn \\
& & + i \,A_{10} \,\smn \,\ga \,S_{\mu} p_{\nu} 
+ i \,{A_{11} \over M^2} \,p \cdot S \ \smn \,\ga \,p_{\mu} P_{\nu},
\eea
where the amplitudes $A_i$ depend on $\sigma \equiv 2p\cdot P$ and
$\tau \equiv p^2$. Hermiticity requires all amplitudes $A_i (\sig, \tau)$ to
be real. Time reversal invariance has also been used.

As the variation along $n_-$ up to order $M^2/Q^2$ is irrelevant in a hard 
process, one encounters the quantities
\be
\Phi^{[\Gam]} \left( x, \ptns \right) = 
\left. \half \int dp^- \ \tr \left( \Phi \Gam \right) \right|_{p^+=xP^+}.
\label{DiracProjPhi}
\ee

The projections of $\Phi$ on different Dirac structures 
define distribution functions. They are related to integrals over linear
combinations of the amplitudes. 
The projections which contribute at leading order are
\begin{eqnarray}
\Phi^{[\gamma^+]}(x,\tv{p}) &\equiv& f_1(x,\tv{p}^2), \nn \\
\Phi^{[\gamma^+\gamma_5]}(x,\tv{p}) &\equiv&
\lambda g_{1L}(x,\tv{p}^2)+\frac{\ptinst}{M}g_{1T}(x,\tv{p}^2), \nn \\ 
\Phi^{[i\sigma^{i+}\gamma_5]}(x,\tv{p}) &\equiv&
S_T^i h_{1T}(x,\tv{p}^2)
+\frac{p_T^i}{M}\left(\lambda h_{1L}^\perp(x,\tv{p}^2)
+\frac{\ptinst}{M}h_{1T}^\perp(x,\tv{p}^2)\right)\nn
\end{eqnarray}

After integration over the quark transverse momentum, these
leading twist functions have well known probabilistic interpretations.
The function $f_1(x)$ gives the probability of finding a quark with light-cone 
momentum fraction $x$ in the $+$ direction (and any transverse momentum).
$g_1(x)$ is a chirality distribution: in a hadron that is in a positive
helicity eigenstate ($\lambda = 1$), it measures the probability of finding a 
right-handed quark with light-cone momentum fraction $x$ minus the
the probability of finding a left-handed quark with the same light-cone
momentum fraction (and any transverse momentum). $h_1(x)$ is a transverse
spin distribution: in a transversely polarized hadron, it measures the
probability of finding quarks
with light-cone momentum fraction $x$ polarized along the direction of
the polarization of the hadron minus the probability of finding quarks
with the same light-cone momentum fraction polarized along the direction
opposite to the polarization of the hadron. 

The subleading projections can also be expressed in distributions.
The integrated distributions are $e(x)$, $\gtx$ and $\hlx$. Important 
relations are
\bea \label{piet1}
\gtx & = & \gox + {d \over dx} \ \gotox, \\
\hlx & = & \hox - {d \over dx} \ \holperpox, \label{piet2},
\eea
where
\be
F^{(n)}(x) \equiv \int d^2 \ptns \ \left( {\pts \over 2M^2} \right)^n
F(x,\pts).
\ee
The functions $\gtwo = \gt - \go $ and $\htwo = 2(\hl - \ho) $
thus satisfy the sum rules
\bea \label{bcsr}
\int_0^1 dx \ \gtwox & = & - \goto (0), \\
\int_0^1 dx \ \htwox & = & 2\,\holperpo (0),
\eea
If the functions $\goto$ and $\holperpo$ vanish at the origin, we
rediscover the Burkhardt-Cottingham
sum rule \cite{bc70} and the Burkardt sum rule \cite{burkardt93}.


\section{The Diquark Spectator Model}

The basic idea of the spectator model is to treat the intermediate 
states that can be inserted in $\Phi$ in Eq.~\ref{qcf1}
as a state with a definite mass $M_R$, called the spectator. 
In other words, we make a specific ansatz for the spectral 
decomposition of the correlation function. 
The quantum numbers of the intermediate state are those determined by 
the action of the quark field on the state $\vert P,S\rangle$, 
hence the name {\em diquark spectator}. 
The inclusion of antiquark and gluon distributions requires
a more complex spectral decomposition of intermediate states. 
We restrict ourselves to the simplest case. $\Phi$ is then given by
\begin{equation}
\label{qcfdq1}
\Phi^{R}_{ij} = \frac{1}{(2 \pi)^3} \;
\langle P,S | \bpsi_j(0) | X^{(\lambda)} \rangle \;
\theta ( P_R^+ )\;\delta \left[ (p-P)^2 - M_{R}^2 \right] \;
\langle X^{(\lambda)} | \psi_i(0) | P,S \rangle
\end{equation}
where $P_R$ = $P-p$ and $X^{(\lambda)}$ represents the spectator, whose
possible spin states are indicated with $\lambda$. 

The spin of the diquark state can be 0 (scalar diquark $s$) or 1 (axial
vector diquark $a$). 
The matrix element appearing in the RHS of (\ref{qcfdq1}) is given by
\bea
\langle X_s | \psi_i(0) | P, S \rangle & = & 
{\left( {i \over \sla p -m } \right)}_{ik} \;\Upsilon^s_{kl} \ U_l(P,S), \\
\langle X_a^{(\lam)} | \psi_i(0) | P, S \rangle & = & \eps_\mu^{*(\lam)}
{\left( {i \over \sla p -m } \right)}_{ik} \;\Upsilon^{a\mu}_{kl} \ U_l(P,S),
\eea
for a scalar and a vector diquark, respectively.
The matrix elements consist of a nucleon-quark-diquark 
vertex $\Upsilon$ yet to be specified, the Dirac spinor
for the nucleon $U_l(P,S)$, a quark propagator for the untruncated quark line
($m$ is the mass of the quark) and a polarization vector $\eps_\mu^{*(\lam)}$
in the case of an axial vector diquark. The next step is to fix the Dirac
structure of the vertex $\Upsilon$. We assume that
\bea
&&\Upsilon^s = {\bf 1} \ g_s(p^2),
\\ && \Upsilon_\mu^a 
=\frac{g_a(p^2)}{\sqrt{3}}\,\gamma_5\,\left(\gamma^\mu + \frac{P^\mu}{M}\right),
\eea
where $g_R(p^2)$ (with $R$ being $s$ or $a$) are form factors that take into 
account the composite structure of the nucleon and the diquark.
In the choice of vertices, the factors and projection operators are
chosen to assure that in the target restframe the vector diquark states
are pure spatial and the nucleon spinors have only upper components, in
which case e.g. the axial vector diquark vertex reduces to $\chi_N^\dagger
\bm \sigma \cdot \bm \epsilon \chi_q$.
With our choices we obtain
\begin{equation}
\label{qcfdq2}
\Phi^R = {\vert g^R(p^2)\vert^2 \over 2(2 \pi)^3} \
{\del \left(\tau-\sigma+ M^2 -M_R^2 \right)
\over {\left( p^2 -m^2 \right)}^2} \,(\sla p + m)\,\left(\sla P + M \right)
\,(1 +a_R \ga \sla S) \ (\sla p +m),
\end{equation}
where $a_s=1$ and $a_a=-1/3$. In obtainig this result we used as the
polarization sum for the axial vector diquark 
$\sum_\lambda\eps_\mu^{*(\lam)} \eps_\nu^{(\lam)}=-g_{\mu\nu}+P_\mu P_\nu/M^2$,
which is consistent with the choice that the axial vector diquark spin states
are purely spatial in the nucleon restframe.
We will use the same form factors for scalar and axial vector diquark being
of the form
\be
\label{ff}
g(\tau) = N\,\frac{\tau - m^2}{\left| \tau - \Lam^2 \right|^\alpha}.
\ee
$\Lam$ is another parameter of the model and $N$ is a
normalization constant. This form factor has the
advantage of killing the pole of the quark propagator \cite{thomas94}.


\section{Results and Discussion}

Using Eq.~\ref{qcfdq2} we can
compute the amplitudes $A_i$ shown in Eq.~\ref{cf2}.
Taking out some common factors by defining
\be
A_i = {N^2 \over 2(2\pi)^3} \
{\del \left( \tau - \sig +M^2 - M_R^2 \right) \over
\left| \tau - \Lam^2 \right|^{2\alpha}} \ \tilde{A}_i
\ee
we obtain $\tilde A_4$ = $\tilde A_5$ = $\tilde A_{12}$ = 0 and
\bea \label{a1}
\tilde{A}_1 & = & - \tilde{A}_6/a_R = 
{m \over M}\,\left( (M+m)^2 - M_R^2\right) +
(\tau - m^2) \left( 1 + {m \over M} \right), \nn \\
\tilde{A}_2 & = & - \tilde{A}_9/a_R = - \left( \tau - m^2 \right), \nn \\
\tilde{A}_3 & = & - \tilde{A}_{10}/a_R = (M+m)^2 - M_R^2 + (\tau-m^2), \nn \\
\tilde{A}_7 & = & 2\,a_R\, mM, \nn \\
\tilde{A}_8 & = & - \tilde{A}_{11}/a_R = 2\,a_R\,M^2. \nn
\eea
Introducing the function
\be
\lambda_R^2(x) = \Lambda^2(1-x) + xM_R^2 - x(1-x)M^2,
\ee
one gets, e.g., for $\foxp$ the result
\be
\foxp = {N^2 \,(1-x)^{2\alpha - 1} \over 16 \pi^3} \ 
{ {\left( xM + m \right)}^2 + \pts \over
\left( \pts + \lambda_R^2 \right)^{2\alpha}}.
\ee

Although there is a certain freedom in the choice of the
parameters, one immediately sees that the occurence of singularities in
the integration region will cause problems that can be avoided if we require
$\lambda_R^2(x)$ to be positive, what imposes the condition $M_R > M-\Lam$.
Provided this condition is fulfilled one obtains the integrated distributions:
\bea
\fox & = & {N^2 \,(1-x)^{2 \alp -1} \over 32 \pi^2\,(\alpha-1)(2\alpha-1)} 
\ \left[ {2(\alpha-1)\,( xM + m)^2 + \lambda_R^2(x) \over 
\left(\lambda_R^2(x)\right)^{2\alpha - 1}} \right], \nn \\
\gox & = & {N^2 a_R \,(1-x)^{2 \alp -1} \over 32 \pi^2\,(\alpha-1)(2\alpha-1)} 
\ \left[ {2(\alpha-1)\,( xM + m)^2 - \lambda_R^2(x) \over 
\left(\lambda_R^2(x)\right)^{2\alpha - 1}} \right], \nn \\
\hox & = & {N^2 a_R \,(1-x)^{2 \alp -1} \over 16 \pi^2 (2 \alp -1)} \
{(Mx + m)^2 \over \left(\lambda_R^2(x)\right)^{2\alpha - 1}}. \nn
\eea
Explicit expressions for the twist three functions can also be written
\cite{joao97}.

An example of a $\bm p_T^2/2M^2$-weighted distribution is
\be
\gotox = {N^2 a_R \,(1-x)^{2 \alp-1} \over 32 \pi^2\,(\alp -1) (2\alpha-1)}
\ {(xM + m)M \over \left(\lambda_R^2(x)\right)^{2\alpha - 2}}.
\ee

Up to now, we have not specified flavor in the distributions. For the 
nucleon we distinguish two types of distributions, $f_1^s$ and 
$f_1^a$, etc. This corresponds to the spectator having spin 0 or spin 1, 
respectively. 
In that case spin 0 diquarks are in a flavor singlet
state (S) and spin 1 diquarks are in a flavor triplet state (T) in order to 
combine to a symmetric spin-flavour wave function.
Since the coupling of the spin has already been included in the vertices, 
we need the flavor coupling
\be
\label{flavour}
|p \rangle = {1 \over \sqrt{2}}\ \vert u \ S_0\rangle +
{1 \over \sqrt{6}} \vert u \ T_0\rangle
- {1\over \sqrt{3}} \ \vert d \ T_1\rangle,
\ee
to find that the flavor distributions are
$f_1^u = \frac{3}{2}\,f_1^s + \frac{1}{2}\,f_1^a$,
$f_1^d = f_1^a$ 
and similarly for $g_1$, $h_1$, $e$, $g_T$ and $h_L$.
The proportionality of the numbers is obtained from Eq.~\ref{flavour}, 
while the overall factor is chosen to reproduce the sum rules for the number 
of up and down quarks if $f_1^s$ and $f_1^a$ are normalized to unity upon 
integration over $\bm p_T$ and $x$. This will fix the normalization $N$ 
in the form factor. The factors $a_s = 1$ and $a_a = -1/3$ in the distribution 
functions will produce different $u$ and $d$ weighting for unpolarized 
and polarized distributions. 
Further differences between $u$ and $d$ distributions can also be induced 
by different choices of $M_R$, $\Lambda$ or $\alpha$. 
We take for the nucleon $\alpha=2$ to reproduce the right large $x$ behavior 
of $f_1^u$, i.e.~$(1-x)^3$, as predicted by the Drell-Yan-West relation and 
reasonably well confirmed by data. 
We refrain from tuning the large $x$ behaviour of $f_1^d$ to match the 
$(1-x)^4$ form indicated by data. Since $f_1^d$ is only
affected by vector diquarks, this could be easily obtained by choosing a
different form factor for the latter. We feel that this kind of fine-tuning 
would take things too far with the simple model we use. 
Similarly, we will only consider one value of $\Lambda$.
We will, however, allow for different masses for scalar and vector diquark
spectators. The color magnetic hyperfine interaction, responsible 
for the nucleon-delta mass difference, will also produce a 
mass difference between singlet and triplet diquark states. Neglecting 
dynamical effects, group-theoretical factors lead to a difference 
$M_a - M_s$ = 200 MeV. 

Another important constraint comes from the axial charge of the nucleon
\be \label{axialcharge}
g_A = \int_0^1 dx \left[ g_1^u(x) - g_1^d(x) \right]
= \int_0^1 dx \left[ {3 \over 2} \ g_1^s(x)
+ {1 \over 2} \ g_1^a(x) \right].
\ee

The sensitivity to the parameters $M_R$ and $\Lambda$ is best illustrated 
by considering some characteristic values.
We take a quark mass of 0.36 GeV, two different values
for $M_R$ (0.6 and 0.8 GeV) and three values for $\Lam$ (0.4, 0.5 and 0.6 GeV.)
In Table~\ref{sacomp} the values of some moments are given.
\begin{table}[h]
\caption{ \label{sacomp}
The second moment of $f_1$, $\langle x \rangle$ =
$\int dx\,xf_1(x)$ and the first moments $g_1$ = $\int dx\,g_1(x)$
and $h_1$ = $\int dx\, h_1(x)$ are given for two diquark masses
and for three values of $\Lambda$.}
\begin{tabular}{c|ccc|ccc}
& \multicolumn{3}{c}{$M_R = 0.6$ GeV} & \multicolumn{3}{c}{$M_R = 0.8$ GeV} \\
$\Lambda$ (GeV) & $\langle x\rangle^R$ & $g_1^R$ & $h_1^R$
& $\langle x\rangle^R$ & $g_1^R$ & $h_1^R$ \\
\hline
0.4 & 0.367 & 0.922$\,a_R$ & 0.965$\,a_R$
& 0.232 & 0.643$\,a_R$ & 0.821$\,a_R$ \\
0.5 & 0.373 & 0.792$\,a_R$ & 0.896$\,a_R$
& 0.255 & 0.524$\,a_R$ & 0.760$\,a_R$ \\
0.6 & 0.383 & 0.664$\,a_R$ & 0.832$\,a_R$
& 0.278 & 0.412$\,a_R$ & 0.707$\,a_R$
\end{tabular}
\end{table}

\begin{figure}
\begin{center}
\psfig{file=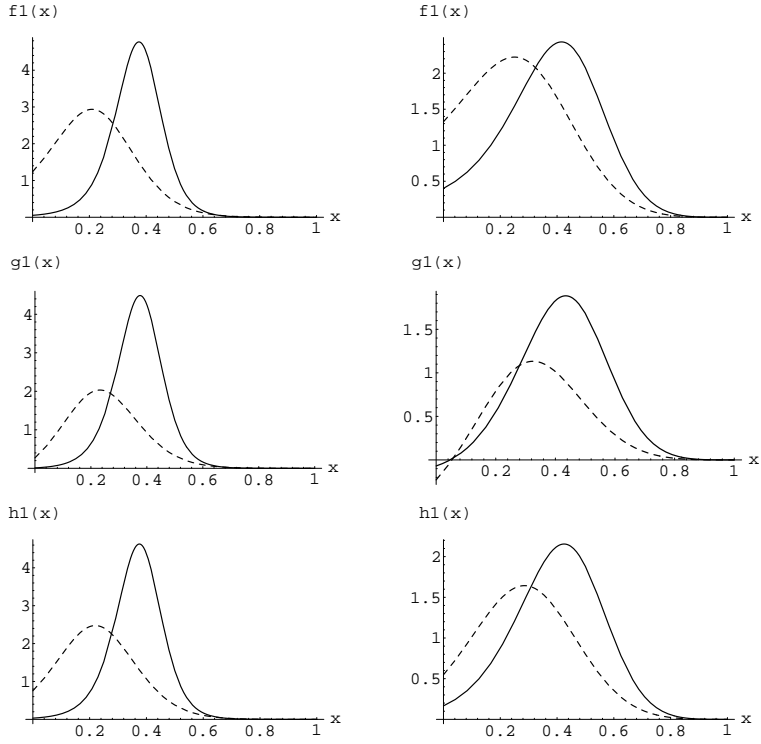,height=15cm}
\vspace{-2cm}
\caption{Twist two distributions for the nucleon.
The plots on the top represent $\fox$, the ones on the middle show $\gox / a_R$ 
and at the bottom we have $\hox /a_R$. 
The plots on the left correspond to $\Lam = 0.4$ GeV 
and the ones on the right to $\Lam = 0.6$ GeV.
The full line corresponds to $M_R = 0.6$ GeV and the dashed line to 
$M_R = 0.8$ GeV.
\label{grtorino3}}
\end{center}
\end{figure}

Fig. \ref{grtorino3} shows the twist two distributions for different
values of the mass of the spectator and of the parameter $\Lam$. 
We can see that an increase of $M_R$ induces a shift on the peak of the 
valence distribution
$\fox$ towards lower values of $x$ and a decrease in its second moment.
This is due to the fact that a spectator with
higher mass corresponds to more massive intermediate states
which can contain sea quarks. In these circumstances the momentum 
carried by valence quarks decreases.
In the non relativistic limit the equality $\fox = \gox = \hox$
holds, a result that could be anticipated. 

%
%

\begin{figure}
\begin{center}
\vspace{-5cm}
\psfig{file=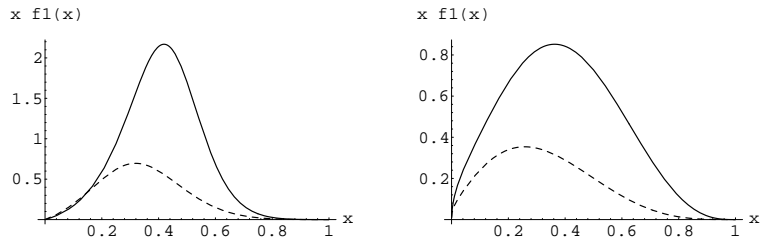,height=15cm}
\vspace{-5cm}
\caption{Twist two distributions for the nucleon.
The plot on the left shows $x f_u^s(x)$ (full line) and $x f_1^d(x)$ 
(dashed line) for $M_s=0.6$ GeV, $M_a = 0.8$ GeV and $\Lam=0.5$ GeV.
The plot on the right shows the ``low scale'' valence distributions of GRV.}
\label{grtorino4a}
\end{center}
\end{figure}

We now use the axial charge of the nucleon to find the most suitable
values for $\Lam$ and the diquark masses. 
The values $\Lam = 0.5$ GeV, $M_s = 0.6$ GeV and $M_a = 0.8$ GeV
give $g_A = 1.25$, close to the experimental value.
Fig. \ref{grtorino4a} shows the distribution $\fox$ multiplied by $x$
for these values of the parameters.
We find a satisfactory qualitative agreement with the valence 
distributions of Gl\"{u}ck, Reya and Vogt \cite{grv95}.

%
%

\begin{figure}
\begin{center}
\vspace{-3cm}
\psfig{file=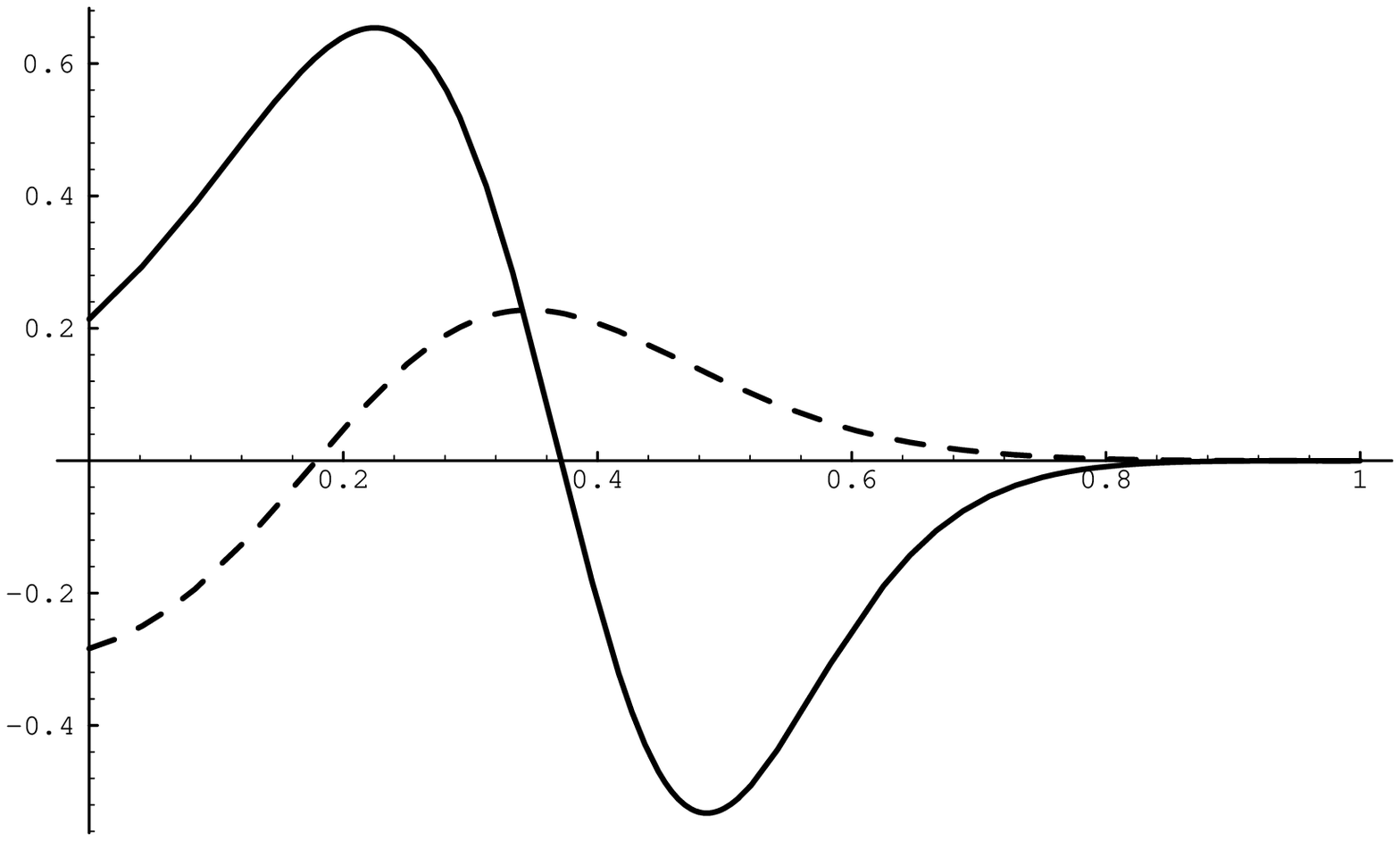,height=10cm}
\vspace{-3cm}
\caption{Distributions $\gtwo^u(x)$ (solid line) and 
$\gtwo^d(x)$ (dashed line) 
for $M_s=0.6$ GeV, $M_a = 0.8$ GeV and $\Lam=0.5$ GeV.
\label{grtorino10}}
\end{center}
\end{figure}

Finally, Fig. \ref{grtorino10} shows the distributions $\gtwo^u(x)$ and 
$\gtwo^d(x)$, for which we find a small violation of the 
Burkhardt-Cottingham sum rule, in agreement with Eq. \ref{bcsr}.

This work was suported by the Foundation for Fundamental Research on Matter
(FOM), the National Organization for Scientific Research and the Junta
Nacional de Investigacao Cientifica (JNICT, PRAXIS XXI).

%
%


\vfill

\end{document}